\documentstyle[twoside,fleqn,espcrc2,epsf]{article}

\newcommand{\lesssim}{ {\
\lower-1.2pt\vbox{\hbox{\rlap{$<$}\lower5pt\vbox{\hbox{$\sim$}}}}\ }  }
\newcommand{\gtrsim}{ {\
\lower-1.2pt\vbox{\hbox{\rlap{$>$}\lower5pt\vbox{\hbox{$\sim$}}}}\ }  }

\newcommand{\be}{\begin{equation}}
\newcommand{\ee}{\end{equation}}
\newcommand{\bea}{\begin{eqnarray}}
\newcommand{\eea}{\end{eqnarray}}
\newcommand{\noi}{\noindent}
\newcommand{\nn}{\nonumber}

\newcommand{\cF}{{\cal F}}

\newcommand{\cL}{{\cal L}}

\newcommand{\cO}{{\cal O}}

\newcommand{\Imm}{\mbox{\rm Im}}
\newcommand{\QCD}{\mbox{\rm {\tiny QCD}}}

\newcommand{\MeV}{\mbox{\rm MeV}}
\newcommand{\GeV}{\mbox{\rm GeV}}
\newcommand{\fm}{\mbox{\rm fm}}
\newcommand{\MS}{\mbox{\rm MS}}

\newcommand{\AmS}{{\protect\the\textfont2
  A\kern-.1667em\lower.5ex\hbox{M}\kern-.125emS}}

\title{Light Quark Masses in QCD}

\author{Eduardo de Rafael\address{Centre de Physique Th\'eorique,  
CNRS-Luminy, Case 907, F-13288 Marseille Cedex 9, France}
        \thanks{Invited talk given at the conference ``QCD 97", Montpellier,
France, July 1997.}}
      
\begin{document}

\begin{abstract}
I report on recent work done in collaboration with Laurent Lellouch and Josep
Taron~\cite{LRT97} concerning the derivation of rigorous lower bounds for the
combination of light quark masses $m_s +m_u$ and $m_d +m_u$.
\end{abstract}

\maketitle

\section{INTRODUCTION}

The light  $u$,  $d$ and $s$ quarks in the QCD Lagrangian have bilinear
couplings
\be
\label{lagrangian}
\cL_{\QCD}=\cL_{\QCD}^{(0)}-\sum_{q=u,d,s}m_{q}{\bar q}(x)q(x)\,,
\ee
with masses $m_q\neq 0$ which explicitly break the chiral symmetry of the
Lagrangian. Twenty five years of efforts of many people (including
significant contributions from the chairman of this Conference
Stephan Narison,) have resulted in the following
values~\cite{BPR95}:
\be
\label{eq:mumdval}
m_{u}(4\GeV^2)+m_{d}(4\GeV^2)=(9.5\pm 1.9)\MeV\,,
\ee
and~\cite{JM95,CDPS95}
\be
\label{eq:mumsval}
m_{u}(4\GeV^2)+m_{s}(4\GeV^2)=(149\pm 25)\MeV\,.
\ee
These are values of the running masses at a scale $\mu^2 = 4\GeV^2$ in the
$\overline{\MS}$ renormalization scheme. [The choice of scale is of course
arbitrary; $4\GeV^2$ has become the conventional choice of the
lattice community. We shall adopt it here as well so as to make easier the
comparison of our bounds with their results.] There are quite a few
independent determinations, including some from numerical simulations of
lattice QCD~\cite{Alletal94}, which now agree  within errors with these values.

The values in (\ref{eq:mumdval}) and (\ref{eq:mumsval}) are smaller than
those obtained in the very early determinations of the light quark masses
using QCD sum rules (see e.g. ref.~\cite{BNRY81} and references therein.)
This is partly due to the fact that
the experimental determination of $\Lambda_{\overline{\MS}}$ has resulted in
larger values than those quoted in the early 80's, and partly to the
appearance of rather large coefficients in the perturbative expansion of the
relevant QCD two--point functions.

Since the summer'96 there have appeared some determinations of
the light quark masses from new analyses of numerical simulations of
lattice QCD~\cite{BG96,Mactal96} which find substantially lower values than
the ones quoted above. More recently, an independent lattice QCD
determination~\cite{Eietal97} using dynamical Wilson fermions finds results
which, for $m_{u}+m_{d}$ are in agreement within errors with those of
refs.~\cite{BG96,Mactal96}; while for
$m_{s}$ they agree rather well, again within errors, with the value quoted in
(\ref{eq:mumsval}). The present situation concerning the values of the light
quark masses has thus become rather confusing again, and this is why we have
decided to concentrate on {\it what can be said on rigorous grounds} about
the light quark mass values at present, rather than argue in favour or
against the reliability of a particular determination.

We shall show that there exist rigorous lower bounds on how small the light
quark masses $m_{s}+m_{u}$ and $m_{d}+m_{u}$ can be. The derivation of the
bounds is based on first principles: the fact that
two--point functions
\be
\label{eq:2pf}
\Psi (q^2)\equiv i\int d^{4}x e^{iq\cdot x}\langle 0\vert T\left(\cO (x)\cO
(0)^{\dagger}\right)\vert 0\rangle\,
\ee 
of local operators $\cO(x)$ obey dispersion relations, and the positivity
of the corresponding hadronic spectral function
\bea
\label{eq:spectral}
 &&\frac{1}{\pi}\mbox{Im}\Psi(t)= (2\pi)^3\ 
\sum_{\Gamma}\,\delta^{(4)}(q-\sum p_{\Gamma})\,\times\nn\\
&&\qquad\times\langle 0\vert \cO(0)\vert\Gamma
\rangle \langle \Gamma\vert \cO^\dagger(0)\vert 0\rangle \,,
\eea
where the sum over $\Gamma$  is extended to all possible  on--shell hadronic
states with the quantum numbers of the operator $\cO$.

We shall be concerned with two types of two--point functions, one where $\cO$
is the divergence of the strangeness changing axial current
\be
\label{eq:dac}
\cO (x)\equiv\partial_{\mu}A^{\mu}(x)=(m_{s}+m_{u}):\!\!{\bar
s}(x)i\gamma_{5}u(x)\!\!:\,;
\ee 
and another one where $\cO$ is the scalar operator $S(x)$, defined as the
isosinglet component of the mass term in the QCD Lagrangian
\be
\label{eq:scc} 
\cO (x)\equiv S(x)=\hat{m}[:\!\!{\bar u}(x)u(x)\!\!:+:\!\!{\bar
d}(x)d(x)\!\!:]\,,
\ee 
with
\be
\hat{m}\equiv\frac{1}{2}(m_u + m_d)\,.
\ee 
We  shall refer to the corresponding two--point functions respectively 
as the ``pseudoscalar channel'' and the ``scalar channel''.
Both two--point functions obey dispersion relations which in QCD
require two subtractions, and it is therefore appropriate to consider their
second derivatives ($Q^2\equiv -q^2$):
\be
\label{eq:disp}
\Psi''(Q^2)=\int_{0}^{\infty} dt\frac{2}{(t+Q^2)^3}\frac{1}{\pi}\Imm\Psi(t)\,.
\ee
The bounds follow from the restriction of the sum over all pos\-sible hadronic
states which can con\-tri\-bute to the spectral function to the state(s) with
the lowest invariant mass. It turns out that, for the two operators in
(\ref{eq:dac}) and (\ref{eq:scc}), the\-se hadronic contributions are well
known phenomenologically; either from experiment or from chiral perturbation
theory ($\chi$PT) calculations. On the QCD side of the dispersion relation,
the two--point functions in question where the quark masses appear as an
overall factor, are known in the deep euclidean region where
$Q^2\gg\Lambda_{\QCD}^2$ from perturbative QCD (pQCD) at the four loop level.
The leading non--perturbative
$\frac{1}{Q^2}$--power corrections which appear in the operator product
expansion when the time ordered product in (\ref{eq:2pf}) is evaluated in the
physical vacuum~\cite{SVZ79} are also known. As we shall see, the bounds we
find cast serious doubts on the reliability of the recent lattice
determinations reported in refs.~\cite{BG96,Mactal96,Eietal97}.

\section{PSEUDOSCALAR CHANNEL}

We shall call $\Psi_{5}(q^2)$ the two--point function in (\ref{eq:2pf})
associated to the divergence of the strange\-ness chan\-ging a\-xial current
in (\ref{eq:dac}). The lowest hadronic state which contributes to the spectral
function in (\ref{eq:spectral}) is the $K$--pole. From eq.~(\ref{eq:disp}) we
then have
\bea
\label{eq:psi5}
\lefteqn{\Psi_{5}''(Q^2)=\frac{2}{(M_{K}^2 +Q^2)^3}2f_{K}^2 M_{K}^4 +}
\nn \\ & & 
\int_{t_{0}}^{\infty} dt \frac{2}{t+Q^2)^3}\frac{1}{\pi}\mbox{Im}\Psi_{5}(t)
\,,
\eea
where $t_{0}=(M_{K}+2m_{\pi})^2$ is the threshold continuum.

It is convenient to introduce the moments $\Sigma_{N}(Q^2)$ of the
hadronic continuum integral
\bea
\label{eq:contint}
\lefteqn{\Sigma_{N}(Q^2)=\int_{t_{0}}^{\infty}dt\frac{2}{(t+Q^2)^3}\times}
\nn \\ & & 
\left(\frac{t_0 +Q^2}{t+Q^2}\right)^{N}\frac{1}{\pi}\Imm\Psi_{5}(t)\,.
\eea
One is then confronted with a typical moment problem (see e.g.
ref.~\cite{AhK62}.) The positivity of the continuum spectral function
$\frac{1}{\pi}\Imm\Psi_{5}(t)$ constrains the moments
$\Sigma_{N}(Q^2)$ and hence the l.h.s. of (\ref{eq:psi5}) where the light
quark masses appear. The most general constraints among the first three
moments for
$N=0,1,2$ are:
\be
\Sigma_{0}(Q^2)\ge 0,\quad \Sigma_{1}(Q^2)\ge 0,\quad
\Sigma_{2}(Q^2)\ge 0\,;
\ee
\be
\label{eq:diff}
\Sigma_{0}(Q^2)-\Sigma_{1}(Q^2)\ge 0,\,,
\Sigma_{1}(Q^2)-\Sigma_{2}(Q^2)\ge 0\,;
\ee
\be
\label{eq:quad}
\Sigma_{0}(Q^2)\Sigma_{2}(Q^2)-\left(\Sigma_{1}(Q^2)\right)^2\ge 0\,.
\ee 
The inequalities in eq.~(\ref{eq:diff}) are in fact trivial unless $2Q^2<
t_{0}$, which constraints the region in $Q^2$ to too small values for pQCD to
be applicable. The other inequalities lead however to interesting bounds
which we next discuss.

\subsection{Bounds from $\Sigma_{0}(Q^2)\geq 0$.}

\noi 
The inequality $\Sigma_{0}(Q^2)\ge 0$ results in a first bound on the
running masses:
\bea
\label{eq:1stbound}
\lefteqn{\left[m_{s}(Q^2)+m_{u}(Q^2)\right]^2 \ge \frac{16\pi^2}{N_c}
\frac{2f_{K}^2 M_{K}^4}{Q^4}\times} \nn \\ & &
\frac{1}{\left(1+\frac{M_{K}^2}{Q^2}\right)^3}
\frac{1}{\left[1+
\frac{11}{3}\frac{\alpha_{s}(Q^2)}{\pi} +\cdots\right]}\,.
\eea 
Notice that this bound is non--trivial in the large--$N_c$ limit
($f_{K}^2\sim\cO(N_c)$) and in the chiral limit ($m_{s}\sim M_{K}^2$). The
bound is of course a function of the choice of the euclidean $Q$--value at
which the r.h.s. in eq.~(\ref{eq:1stbound}) is evaluated. For the bound to be
meaningful, the choice of $Q[\GeV]$ has to be made sufficiently large so that
a pQCD evaluation of the QCD factor
\be
\label{eq:qcdf0}
\cF_{0}^{\QCD}(Q^2)=1+\frac{11}{3}\frac{\alpha_{s}(Q^2)}{\pi}+\cdots\,,
\ee 
associated to the two--point function $\Psi''_{5}$ is applicable. The dots
in (\ref{eq:qcdf0})
represent higher order terms which have been calculated up to
${\cal O}(\alpha_s^3)$~\cite{C96}, as well as  non--perturbative power
corrections of
$\cO\left(\frac{1}{Q^4}\right)$~\cite{BNRY81,PR82}, and strange quark mass
corrections of
$\cO\left(\frac{m_{s}^2}{Q^2}\right)$ and
$\cO\left(\frac{m_{s}^4}{Q^4}\right)$ including ${\cal O}(\alpha_s)$
terms~\cite{Ge90,SCh88,ChGS85,JM95}. We consider
that,
$Q\gtrsim 1.4\:\GeV$ is a safe choice for a numerical estimate of the
bounds: the evaluation of
$\cF_{0}^{\QCD}$ at
$Q=1.4\:\GeV$ with or without inclusion of the ${\cal O}(\alpha_s^3)$ terms
differs by less than 10\%. 

The lower bound which follows from
eq.~(\ref{eq:1stbound}) for
$m_u + m_s$ at a renormalization scale
$\mu^2=4\:\GeV^2$ results in the solid curves shown in Fig.~1 below. 
\begin{figure}[htb]
\centerline{\epsfbox{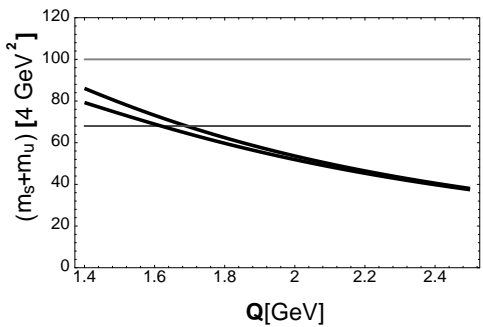}}
\vspace*{-0.7cm}
\caption{\protect\small\it Lower bound in {\rm MeV} for
$[m_{s}+m_{u}](4\:\protect\GeV^2)$ versus $Q(\protect\GeV)$ from
eq.~(\protect\ref{eq:1stbound}).}
\vspace*{-0.5cm}
\label{fig:msmu1}
\end{figure}
\noi
These are the lower bounds obtained by letting
$\Lambda^{(3)}_{\overline{MS}}$ vary~\cite{PDB} between 290~MeV 
(the upper curve) and 380~Mev (the lower
curve) and using $f_{K}=113\:\MeV$, $M_{K}=493.67\:\MeV$. We take the number
of active flavours
$n_f=3$ in all the numerical analyses of this work. Notice that the hadronic
continuum integral in (\ref{eq:contint}) is always larger than the
contribution from three light flavours; and indeed, we have checked that using
$n_{f}=4$ in the QCD expressions leads to even higher bounds. Values of the
quark masses below the solid curves in Fig.~1 are forbidden by the bounds. For
comparison, the horizontal lines in the figure correspond to the central
values obtained by the authors of ref.~\cite{BG96}: their ``quenched'' result
is the horizontal upper line; their ``unquenched'' result the horizontal
lower line.

We wish to emphasize that in Fig.~1 we are comparing what is meant to be a
``calculation'' of the quark masses --the horizontal lines which are the
lattice results of ref.~\cite{BG96}-- with a bound  which in fact can only be
saturated in the limit where
$\Sigma_{0}(Q^2)= 0$. Physically, this limit corresponds to the extreme case
where the hadronic spectral function from the continuum of states is totally
neglected! What the plots show is that, even in that extreme limit, and  for
values of $Q$ in the range
$1.4\:\GeV\lesssim Q\lesssim 1.8\:\GeV$ the new lattice results are already in
difficulty with the bounds.

\subsection{Bounds from the Quadratic Inequality}

The quadratic inequality in (\ref{eq:quad}) results in improved lower
bounds for the quark masses which we show in Fig.~2 below. 
\begin{figure}[htb]
\centerline{\epsfbox{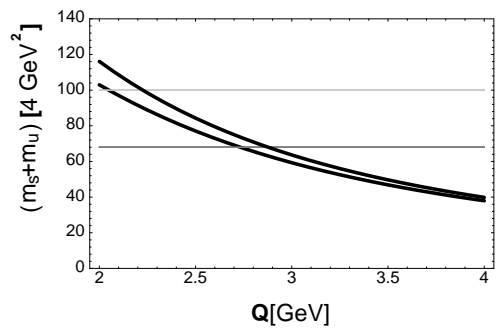}}
\vspace*{-0.7cm}
\caption{\protect\small\it Lower bound in {\rm MeV} for
$[m_{s}+m_{u}](4\:\protect\GeV^2)$ from the quadratic inequality.}
\vspace*{-0.5cm}
\label{fig:msmu2}
\end{figure}
\noi
Like in Fig.~1, these are the lower bounds obtained by letting
$\Lambda^{(3)}_{\overline{MS}}$ vary~\cite{PDB} between 290~MeV (the upper
curve) and 380~Mev (the lower curve). Values of the quark masses below the
solid curves in Fig.~2 are forbidden by the bounds. The horizontal lines in
this figure are also the same lattice results as in Fig.~1.

The quadratic bound is saturated for a $\delta$--like spectral function
re\-pre\-sen\-ta\-tion of the hadronic continuum of states at an arbitrary
position and with an arbitrary weight. This is certainly less restrictive
than the extreme limit with the full hadronic continuum neglected, and it is
therefore not surprising that the quadratic bound happens to be better than
the ones from
$\Sigma_{N}(Q^2)$ for
$N=0,1,$ and $2$. Notice that the quadratic bound in Fig.~2 is plotted at
higher
$Q$--values than the bound in Fig.~1. The evaluation of the corresponding QCD
factor with or without inclusion of the
$\cO(\alpha_{s}^{3})$ terms differs by less than 10\% for
$Q^2\ge 4\:\GeV^2$  and we consider therefore that for the evaluation of the 
quadratic bound $Q\gtrsim 2\:\GeV$ is already a safe choice. We find that
even the quenched lattice results of refs.~\cite{BG96,Mactal96} are in
difficulty with these bounds.

Similar bounds can be obtained for $m_{u}+m_{d}$ when one considers the
two--point function associated to the divergence of the axial current
\be
\partial_{\mu}A^{\mu}(x)=(m_{d}+m_{u}):\!\!{\bar d}(x)i\gamma_{5}u(x)\!\!:\,.
\ee 
The method to derive the bounds is exactly the same as the one discussed
above and therefore we only show, in Fig.~3 below, the results for the
corresponding lower bounds  which we obtain from the quadratic inequality.
\begin{figure}[htb]
\centerline{\epsfbox{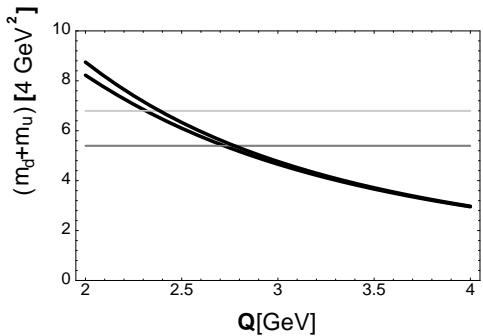}}
\vspace*{-0.7cm}
\caption{\protect\small\it Lower bound in {\rm MeV} for
$[m_{d}+m_{u}](4\:\protect\GeV^2)$ from the quadratic inequality.}
\vspace*{-0.5cm}
\label{fig:mumd1}
\end{figure}
\noi 
We find again that the lattice QCD results of
refs.~\cite{BG96,Mactal96,Eietal97} for $m_{u}+m_{d}$ are in serious
difficulties with these bounds. Notice that in this case the mass corrections
in the QCD two--point function are negligible at the plotted $Q$--values.

\section{SCALAR--ISOSCALAR CHANNEL}

The basic difference between the ``scalar'' and ``pseudoscalar'' channels is
that the lowest state which contributes to the hadronic spectral function in
the ``scalar'' channel is not a pole but the
$\pi-\pi$ continuum with
$J=0$ and $I=0$. This contribution provides a lower bound to the full
spectral function:
\bea
\label{eq:specin}
\lefteqn{\frac{1}{\pi}\Imm\Psi(t)\ge\frac{1}{16\pi^2}\sqrt{1-\frac{4m_{\pi}^2}
{t}}\times} \nn \\
 & & 3\,\vert F(t)\vert^{2}\theta(t-4m_{\pi}^2)\,,
\eea 
where $F(t)$ denotes the scalar--isoscalar pion form factor
\be
\langle \pi^{a}(p)\pi^{b}(p')\vert S(0)\vert 0\rangle =\delta_{ab}
F(t)\,,
\ee
with $t=(p+p')^2$.
In particular, the value of this form factor at the origin is the so
called {\it pion sigma term}. This form factor has been well studied in
$\chi$PT at the one and two loop level (see e.g. refs\cite{DGL90,GM91} and
references therein), and the predicted low energy shape agrees well with the
available experimental information. With
\be
\label{eq:formf} 
F(t)=F(0)\left[1+\frac{1}{6}\langle
r^{2}\rangle_{s}^{\pi}t+
\cO(t^2)\right]\,,
\ee 
it is found~\cite{GM91} that
\be
\label{eq:ff0} 
F(0)=m_{\pi}^{2}\times (0.99\pm 0.02) + \cO(m_{\pi}^{6})\,,
\ee
and 
\be
\langle r^{2}\rangle_{s}^{\pi}= (0.59\pm 0.02)~\fm^{2}\,.
\ee
The pion mass used in eq.~(\ref{eq:ff0}) in ref.~\cite{GM91} is the
$\pi^{+}$--mass. Since the electromagnetic interactions are neglected, we
find it more appropriate to use the
$\pi^{0}$--mass instead, which in any case results in a lower contribution to
the average quark mass. We also wish to point out that in the generalized
version of
$\chi$PT (see e.g. ref.~\cite{KMS95} and references therein,) the value of
$F(0)$ could be sensibly larger, which would result in even larger lower
bounds for the quark masses. A value for the curvature of the scalar form
factor is also quoted in ref.~\cite{GM91}; it involves however some extra
phenomenological assumptions which go beyond
$\chi$PT and this is why we have not included this extra information here.

Using standard methods, (see ref.~\cite{deRT94} and references
therein, and also refs.\cite{AMS75,AEMS75,Mi73}) one can construct bounds on
$\hat{m}$ when something is known about the scalar--isoscalar pion form factor
$F(t)$. For this, it is convenient to map the complex $q^2$--plane to the
complex unit disc:
\be 
i\frac{1+z}{1-z}=\sqrt{\frac{t}{4m_{\pi}^2}-1}\,.
\ee
With $F(0)$ and $F'(0)$ as input, the resulting lower bound for $\hat{m}^2$
is
\bea
\label{eq:bounds}
\lefteqn{\hat{m}^2(Q^2)\ge
\frac{3\pi}{N_c}
\frac{4m_{\pi}^2}{Q^4}\frac{F(0)^{2}}
{\left(\sqrt{1+\frac{4m_{\pi}^2}{Q^2}}+\frac{2m_{\pi}}{Q} \right)^6}\times}
\nn \\ & & 
\frac{1+\left\{-\frac{1}{2} +3z[-Q^2]-
\frac{8m_{\pi}^2}{3}\langle r^{2}\rangle_{s}^{\pi}
\right\}^2}{\left[1+\frac{11}{3}\frac{\alpha_{s}(Q^2)}{\pi}+\cdots
\right]}\,.
\eea 
Notice that this bound is somewhat similar to the {\it first bound}
obtained from the divergence of the axial current in eq.~(\ref{eq:1stbound}).
There are however some interesting differences which we wish to point out.
From the point of view of the large--$N_c$ expansion, the r.h.s. of
eq.~(\ref{eq:bounds}) is
$1/N_c$--suppressed. Also,  from the point of view of the chiral expansion the
r.h.s. of eq.~(\ref{eq:bounds}) has an extra factor $m_{\pi}^2$ [see
eq.~(\ref{eq:ff0})] as compared to the r.h.s. of eq.~(\ref{eq:1stbound}). One
might expect therefore that the lower bound in (\ref{eq:bounds}) should be
much worse than the one in (\ref{eq:1stbound}). Yet the bound for $\hat{m}$ in
eq.(\ref{eq:bounds}) is surprisingly competitive. The reason for this is
twofold. On the one hand, the absence of a prominent narrow resonance in the
$J=0$, $I=0$ channel is compensated by a rather large contribution from the
chiral loops, a phenomenological feature which appears to be common to all
the explored physical processes where the  $J=0$,
$I=0$ channel contributes. On the other hand the fact that, numerically,
$m_{\pi}^2$ and $f_{\pi}^2$ are not so different. The lower bounds for the
sum of running masses
$(m_{u}+m_{d})$ at the renormalization scale $[\mu^2=4\:\GeV^2]$ which follow
from this {\it scalar channel bound} are the dashed curves shown in Fig~4
below
\begin{figure}[htb]
\centerline{\epsfbox{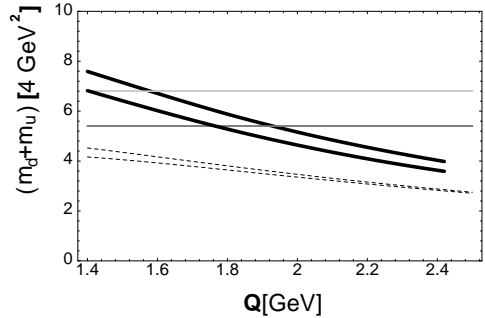}}
\vspace*{-0.7cm}
\caption{\protect\small\it Lower bounds in {\rm MeV} 
from the S\-ca\-lar Chan\-nel for
$[m_{d}+m_{u}](4\:\protect\GeV^2)$.}
\vspace*{-0.5cm}
\label{fig:mumd2}
\end{figure}
\noi
The dashed curves in Fig~4 correspond to the choice
$\Lambda_{\overline{MS}}^{(3)}=290\:\MeV$ (the upper dashed curve) and
$\Lambda_{\overline{MS}}^{(3)}=380\:\MeV$ (the lower dashed curve). The plots
show that for values of $Q$ in the range $1.4\:\GeV\lesssim Q\lesssim
1.6\:\GeV$ the lattice QCD results in~\cite{BG96,Mactal96,Eietal97} are
dangerously close to lower bounds obtained with very little phenomenological
input, (just the values of the scalar form factor at the origin and its
slope at the origin as well.) Notice also that the relevant two--point
function in this channel is practically the same as in the pseudoscalar
channel for which
$Q\gtrsim 1.4\:\GeV$ is already a safe choice.

It is still possible to improve these bounds by taking into account the extra
information provided by the fact that the phase of the scalar form factor
$F(t)$ in the elastic region $4m_{\pi}^2\le t \le 16m_{\pi}^2$ is precisely
the $I=0$, $J=0$
$\pi -\pi$ scattering phase--shift and there is information on this
phase--shift both from $\chi$PT and from experiment. The technology to
incorporate this information is discussed in refs.\cite{AMS75,AEMS75,Mi73}. We
have restricted the phase--shift input to a conservative region: $4m_{\pi}^2
\le t \le (500\:\MeV)^2$, where a resommation of the chiral logarithms 
calculated at the two--loop level in $\chi$PT~\cite{DGL90,GM91} is  certainly
expected to be reliable and is in good agreement with experiment. This
results in improved bounds which correspond to the solid curves shown in
Fig.~4. The two curves  reflect the variation within the quoted errors of the
input parameters. Again, we find that the lattice determinations of
refs.~\cite{BG96,Mactal96,Eietal97}  do not satisfy lower bounds which only
incorporate  very little hadronic input.

\section{CONCLUSIONS}
 
We conclude that the lattice results of refs.~\cite{BG96,Mactal96,Eietal97}
are in serious difficulties with rigorous lower bounds which can be derived
from general properties of QCD and with a minimum of well established
phenomenological input; and this in two different hadronic channels. The
lower bounds we have derived are perfectly compatible with the sum rules
results of refs.~\cite{BPR95,JM95,CDPS95} and earlier references
therein. They are also
compatible with a recent semiempirical determination of $m_{s}$ made by the
ALEPH collaboration~\cite{ALEPH} based on the rate of the hadronic
$\tau$--decays into strange particles observed at LEP. 

Other work with
various claims on possible values of the light quark masses has recently
appeared in refs.~\cite{Co97,Do97,Yn97,DN97}.  

\goodbreak

\noi
{\bf Discussions}

\noi
{\bf C.~Allton}, UKqcd collaboration\\
{\it Lattice theorists use a renormalization scale of $\mu=2\GeV$ since they
don't believe perturbation theory down to $\mu=1\GeV$.}

\noi
{\bf E.~de Rafael}\\
{\it I am aware of that. I understand of course that pQCD is only valid at
sufficiently large euclidean scales; however, once you have pined down the
quark masses at a given scale, you can always use the renormalization group
equation to run them down to whatever conventional scale (larger than
$\Lambda_{\overline{MS}}$) you want.

\vspace*{0.5cm}
\noi
{\bf A.~Pivovarov}, Moscow\\
{\it How worse (or better?) would be your bounds if you switch off QCD
corrections at all?}

\noi
{\bf E.~de Rafael}\\
{\it As I said, the pQCD corrections are large and positive; therefore,
if ignored, the lower bounds for the light quark masses would be
significantly larger.}      


\begin{thebibliography}{99}

\bibitem{LRT97} 
         L. Lellouch, E. de Rafael and J. Taron,  hep-ph/9707523; to appear
         in Phys. Letters {\bf B}

\bibitem{BPR95}
         J.~Bijnens, J.~Prades, and E.~de Rafael, Phys. Lett. {\bf 348B}
         (1995) 226; and references therein.

\bibitem{JM95}
         M.~Jamin and M.~M\"{u}nz, Z. Phys. {\bf C66} (1995) 633; and
         references therein.

\bibitem{CDPS95}
         K.G.~Chetyrkin {\it et al}, Phys.
         Rev. {\bf D51} (1995) 5090; and references therein.

\bibitem{Alletal94}
         C.R.~Allton {\it et al}, Nucl. Phys. {\bf B431} (1994) 667.

\bibitem{BNRY81}
         C.~Becchi {\it et al}, Z. Phys.{\bf C8} (1981) 335.

\bibitem{BG96}
         R.~Gupta and T.~Bhattacharya, Phys. Rev. {\bf D55} (1997) 7203.

\bibitem{Mactal96}
         B.J.~Gough {\it et al}, Phys. Rev. Lett. {\bf 79} (1997) 1622.

\bibitem{Eietal97}
         N.~Ei\-cker {\it et al}, SE\-SAM--Co\-lla\-bo\-ra\-tion,
         pre\-print hep-lat/9704019.

\bibitem{SVZ79}
         M.A.~Shifman, A.I.~ Vainshtein, and V.I.~Zakharov, Nucl. Phys. {\bf
         B147} (1979) 385,448.

\bibitem{S89} 
         L.R.~Surguladze, Sov. J. Nucl. Phys. {\bf 50} (1989) 372.

\bibitem{GKLS90}
         S.G.~Gorishny {\it et al}, Mod.
         Phys. Lett. {\bf A5} (1990) 2703.

\bibitem{GKLS91} 
         S.G.~Gosrishny {\it et al},
         Phys. Rev. {\bf D43} (1991) 1633.

\bibitem{C96} 
         K.G.~Chetyrkin, Phys. Lett. {\bf B390} (1997) 309.

\bibitem{PR82}
         P.~Pascual and E.~de Rafael, Z. Phys. {\bf C12} (1982) 127.

\bibitem{Ge90}
         S.C.~Generalis, J. Phys. G {\bf 16} (1990) 785.

\bibitem{SCh88}
         V.P.~Spiridonov and K.G.~Chetyrkin, Sov. J. Nucl. Phys. {\bf 47}
         (1988) 522.

\bibitem{ChGS85} 
         K.G.~Chetyrkin, S.G.~Go\-rish\-ny, and V.P.~Spi\-ri\-do\-nov, 
         Phys. Lett. {\bf B160} (1985) 149.

\bibitem{AhK62}
         N.I.~Ahiezer and M. Krein, ``Some Questions in the Theory of
         Moments'', Translations of Math. Monographs, 
         Vol.~2; Amer. Math. Soc., Providence, Rhode Island (1962).

\bibitem{PDB}
        {\it Review of Particle Properties}, Phys. Rev. {\bf D54} (1996) 1.

\bibitem{DGL90}
         J.F.~Donoghue, J.~Gasser, and H.~Leutwyler, Nucl. Phys. {\bf B343}
         (1990) 341.

\bibitem{GM91}
         J.~Gasser and U.G.~Meissner, Nucl. Phys. {\bf B357} (1991) 90.

\bibitem{KMS95}
         M.~Knecht and J.~Stern, {\it
         Generalized Chiral perturbation Theory}, in {\bf The Second Daphne
         Physics Handbook, Vol I} (1995), 169.

\bibitem{deRT94}
         E.~de Rafael and J.~Taron, Phys. Rev. {\bf D50}
         (1994) 373.

\bibitem{AMS75}
         G.~Auberson, G.~Mahoux, and F.R.A.~Sim\~ao, Nucl. Phys. {\bf B98}
         (1975) 204.

\bibitem{AEMS75}
         G.~Auberson {\it et al}, Ann. Inst. Henri
         Poincar\'e {\bf XXII} (1975), 317.

\bibitem{Mi73}
         M.~Micu, Phys. Rev. {\bf D7} (1973) 2136.

\bibitem{ALEPH}
         M.~Davier, Nucl. Phys. {\bf B} (Proc. Suppl.) {\bf 55C} (1997), 395.

\bibitem{Co97}
         P.~Colangelo {\it et al}, hep-ph/9704249

\bibitem{Do97}
						   C.A.~Dominguez, hep-ph/9708234.

\bibitem{Yn97} 
         F. Yndur\'ain, hep-ph/9708300.

\bibitem{DN97} 
         H. Dosch and S. Narison, hep-ph/9709215.

\end{thebibliography}
\end{document}